# The power of flexible lattice in $Ca_3Ru_2O_7$:
# Exquisite control of the electrical transport via anisotropic magnetostriction

Hengdi Zhao[1], Hao Zheng[2], Jasminka Terzic[3], Wenhai Song[4], Yifei Ni[1], Yu Zhang[1], Pedro Schlottmann[5] and Gang Cao[1*]

[1]Department of Physics, University of Colorado at Boulder, Boulder, CO 80309, USA
[2] *Argonne National Laboratory, Lemont, IL 60439, USA*
[3] *Department of Physics, Western Kentucky University, Bowling Green, KY 42101, USA*
[4] *Institute of Solid State Physics, Chinese Academy of Sciences, Hefei, China*
[5] *Department of Physics, Florida State University, Tallahassee, FL 32306, USA*

$Ca_3Ru_2O_7$ is a correlated and spin-orbit-coupled system with an extraordinary anisotropy. It is both interesting and unique largely because this material exhibits conflicting phenomena that are often utterly inconsistent with traditional precedents, particularly, the quantum oscillations in the nonmetallic state and colossal magnetoresistivity achieved by avoiding a fully spin-polarized state. This work focuses on the relationship between the lattice and transport properties along each crystalline axis and reveals that application of magnetic field, H, along different crystalline axes readily *stretches* or *shrinks* the lattice in a uniaxial manner, resulting in distinct electronic states. Furthermore, application of modest pressure drastically amplifies the *anisotropic* magnetoelastic effect, leading to either an occurrence of a robust metallic state at H || hard axis or a reentrance of the nonmetallic state at H || easy axis. $Ca_3Ru_2O_7$ presents a rare lattice-dependent magnetotransport mechanism, in which the extraordinary lattice flexibility enables an exquisite control of the electronic state via magnetically stretching or shrinking the crystalline axes, and the spin polarization plays an unconventional role unfavorable for maximizing conductivity. At the heart of the intriguing physics is the anisotropic magnetostriction that leads to exotic states.

*gang.cao@colorado.edu

*Introduction.* Ruthenates host extended *4d*-electron orbitals and comparable energy scales among on-site Coulomb interactions (0.5-3 eV), crystal fields, Hund's rule energies, *p-d* orbital hybridization, and spin-lattice (magnetoelastic) coupling. Spin-orbit interactions (~ 0.16 eV) are weaker than the above but consequential [1]. A combined effect of all these factors renders an extraordinary susceptibility of the ground state to lattice distortions that hallmarks this class of materials. The contrasting ground states of sister compounds $Ca_2RuO_4$ and $Sr_2RuO_4$ is a testament to the decisive role the lattice plays, in which the rotations/tilts of $RuO_6$ octahedra and the lack thereof result in an antiferromagnetic (AFM) insulating state [2] and superconducting state [3], respectively. Indeed, external stimuli that couple to the lattice, such as magnetic field, pressure and electrical current, can readily generate strong and often disproportional responses in structural and physical properties [1].

$Ca_3Ru_2O_7$ [4] is an archetype of this class of materials. It exhibits signatures of most ordered states (except superconductivity) known in condensed matter physics [1, 5-12]. However, what makes it both unique and intriguing are its coexisting contradictory phenomena that are often starkly inconsistent with traditional arguments, and most strikingly, the strong quantum oscillations in the nonmetallic state and colossal magnetoresistance achieved by avoiding a fully polarized-spin state [9-11, 13] (**Fig.1a** [1]).

$Ca_3Ru_2O_7$ undergoes an AFM transition at $T_N$ = 56 K while remaining metallic, and then an abrupt Mott-like transition at $T_{MI}$ = 48 K to a nonmetallic state [4]. The $Ru^{4+}$ ($4d^4$) ions host a low spin S = 1 state because of a large splitting between the $t_{2g}$ and $e_g$ orbitals (> 2 eV) and the relatively modest Hund's rule energy (0.5-0.6 eV). It features a first-order metamagnetic transition, $H_c$, at 6 T, when the magnetic field, H, is applied along the *b* axis, the easy axis; at H ≥ $H_c$, the magnetization along the *b* axis, $M_b$, gets nearly fully polarized, reaching 1.80 $\mu_B$/Ru or 90% of the anticipated saturation moment (2 $\mu_B$/Ru) for an S = 1 state (**Fig.1a**) [4, 10]. A strong anisotropy field of 22.4 T due to the spin-orbit interactions [8, 14] renders a weakly polarized magnetization along the *a* or *c* axis, $M_a$ or $M_c$, (**Fig.1a**). What is remarkably unusual is the corresponding magnetotransport with H applied along the *a*, *b* and *c* axis, respectively. *For* H||*b axis or the easy axis*, the spin polarization expectedly reduces spin scattering, thus the *c*-axis resistivity, $\rho_c$(H||*b*), by one order of magnitude [10]; but, as H further rises, $\rho_c$(H||*b*) linearly increases instead by more than 30% (**Fig.1a**). In contrast, *for H*||*a axis*, the hard axis, $\rho_c$(H||*a*) is reduced by three orders of magnitude (**Fig.1a**) [10]; moreover, *for H*||*c axis*, strong quantum oscillations occur, with $\rho_c$(H||*c*) 10-fold smaller than $\rho_c$(H||*b*) at 45 T (**Fig.1a**) [9, 11]. In short, the strong spin polarization, a hallmark of this ruthenate, leads to the *least reduction of resistivity,* and the colossal magnetoresistance and quantum oscillations are attained only when such a spin-polarized state is avoided. Conventionally, the spin polarization plays an essential role because it optimizes electron hopping by minimizing spin scattering [e.g.,15, 16]. The contradictory behavior in **Fig.1a** (which is also true for the *a*-axis resistivity $\rho_a$ [10]) points to a hidden and yet particularly decisive component, the lattice. The application of relatively low magnetic field enables the switching between the distinct electronic states, which is another silent feature contrasting with that of spin-mediated transitions often facilitated by higher fields (e.g., 100 T or higher) owing to the gap between spin states [e.g., 17-20] (recall 10 T ~ 1 meV). Such unusual transport has thus far escaped adequate attention and remained largely uninvestigated despite more than two decades of extensive studies of $Ca_3Ru_2O_7$ since its discovery [4]. It is encouraging that recent efforts have gained more insights into the ruthenate [12, 21].

The hiatus in the understanding of this complex material has motivated this work that investigates the relationship between the *lattice* and *transport* properties of single-crystal

Ca$_3$Ru$_2$O$_7$ via magnetostriction and resistivity as functions of magnetic field and pressure. Because of the highly anisotropic nature of this material, these measurements are methodically carried out along the *a* and *b* axis, respectively, so that a direct correlation between the lattice and transport properties along each crystalline axis can be established. Indeed, as this work reveals, application of magnetic field along the *a* or *b* axis readily *stretches* or *shrinks* the lattice in a uniaxial manner, resulting in distinct electronic states. In addition, this work also finds that application of modest hydrostatic pressure not only considerably broadens the $t_{2g}$ bandwidth but also drastically amplifies the *anisotropic* magnetoelastic effects, which in turn leads to either an occurrence of a robust metallic state when H∥*a* axis (hard axis) or a reentrance of the nonmetallic state when H∥*b* axis (easy axis). All this outlines a picture of lattice-dependent magnetotransport in which the unusual lattice flexibility enables a control of the electronic state via magnetically stretching or shrinking the *a* or *b* axis, and the spin polarization plays an unconventional role unfavorable for maximizing conductivity. Note that the coupling between magnetostriction and transport/magnetic properties has been reported in materials such as cobaltites and manganites [e.g.,15-20], in which magnetostrictive effects are isotropic, and the spin polarization always remains an essential, favorable element that facilitates either electron hopping [15] or transitions between different spin states [19]. This is not the case in Ca$_3$Ru$_2$O$_7$. The unusually anisotropic magnetostriction is at the heart of the intriguing physics of Ca$_3$Ru$_2$O$_7$, hinting an analogy to multiferroics.

*Results and discussion.* The single crystals were grown using a flux method [5, 22]. The dilatometer is made with four identical strain gauges forming a Wheatstone bridge to cancel any unwanted changes. Copper with a known thermal expansion coefficient [23] was measured to ensure accuracy. See Ref. 24 for more experimental details.

Ca$_3$Ru$_2$O$_7$ adopts an orthorhombic structure with space group *A2$_1$ma* (No.36) and lattice parameters *a* = 5.3720(6) Å, *b* = 5.5305(6) Å, and *c* = 19.572(2) Å at room temperature [25]. Note that the *a* axis is significantly shorter than the *b* axis, and their difference defines the orthorhombicity, |*a-b*|/[1/2(*a+b*)] ~ 3% at ambient conditions. The structure distortions are closely associated with tilts of the RuO$_6$ octahedra [25]. The tilt projects primarily onto the *ac* plane (153.22°, compared to 180°) and only slightly affects the *bc* plane (172.0°) [25]. These bond angles determine the overlap matrix elements between the *t$_{2g}$* orbitals, thus directly impact the band structure [5]. Indeed, density functional calculations indicate a complex coupling of the *d$_{xy}$* sheet with the *d$_{xz}$* and *d$_{yz}$* bands owing to the tilts of RuO$_6$ octahedra, which causes the large electronic anisotropy [8]. Moreover, the *c* axis rapidly contracts by 0.1% below T$_{MI}$ over an interval of 40 K - 48 K (**Fig.1b** [11]), supplementing a spin reorientation and orbital redistribution [10, 26, 27]. Below 40 K, the *c*-axis contraction completes, leading to compressed RuO$_6$ octahedra [27]. The x-ray diffraction data show a weak but discernable anomaly in the *a* axis near T$_N$ but not near T$_{MI}$ (**Fig.1b**). Using a sensitive dilatometer, this work uncovers an abrupt, negative thermal expansion along both the *a* and *b* axis occurring at T$_{MI}$ (**Fig.1c**); the corresponding thermal expansion coefficient α is - 430 ppm/K and - 250 ppm/K for the *a* and *b* axis, respectively. The nonmetallic state is a consequence of these lattice changes that dictate the splitting of *d$_{xy}$*, *d$_{zx}$* and *d$_{yz}$* orbitals. Previous studies indicate an orbital order with a charge gap of ~ 0.1 eV [27, 28, 29] or a weak orbital polarization in the nonmetallic state [30] or polar domains [12]. Nevertheless, the overlap matrix elements between the *t$_{2g}$* orbitals hinge on the orthorhombic distortions stabilized by the tilts/rotations of RuO$_6$ octahedra.

Now let us turn to the magnetostriction, ΔL/L. L is the lattice parameter, either the *a* or *b* axis; ΔL is the field-induced change in the lattice parameter, defined as ΔL = L(H) – L(0). ΔL/L is measured for both H∥L and H⊥L, respectively, whose contrasting behavior reveals an exceptionally anisotropic magnetostriction (**Fig.2**). Two temperature regions are assigned based on the response of the *a* and *b* axis to H.

In Region I (T < 40 K), two pronounced anomalies are observed in both ΔL/L (H∥*a* axis) (**Figs.2a-2d**) and ΔL/L (H∥*b* axis) (**Figs.2A-2D**). A sharp peak at $H_c$ = 6 T corresponds to the first-order metamagnetic transition. As H further increases, another broader yet distinct anomaly, H*, emerges, marking an onset of a rapid decrease in the lattice parameters. Unlike $H_c$, H* is strongly T-dependent, decreasing with increasing T. An important trend is that when H∥*a* axis or ⊥ *b* axis, the *a* axis stretches but the *b* axis shrinks for H < H* (**Figs.2a-2d**). In contrast, when H∥*b* axis or ⊥ *a* axis, the *b* axis expands more significantly whereas the *a* axis shrinks for H<H* (**Figs.2A-2D**).

The magnetostrictive effects are reversed in Region II (40 K ≤ T < 48 K). Both the *a* and *b* axis shrink instead when H∥*a* and *b* axis, respectively. For ΔL/L (H∥*a* axis) (**Fig.2e-2f**), the *a* axis decreases faster than the *b* axis, whereas for ΔL/L (H∥*b* axis) (**Fig.2E-2F**), the *b* axis decreases faster than the *a* axis. Both *a* and *b* axis show an anomaly at H*. The distinct behavior of ΔL/L suggests a lattice environment different from that below 40 K because the *c* axis undergoes a rapid change in Region II (**Fig.1b**), which accompanies the concurrent spin reorientation and orbital redistribution [5, 10, 26, 27, 29].

The data in **Fig.2** also invoke the dependence of the magnetostriction on the magnetization and magnetic easy axis: ΔL/L = (ΔL/L)$_s$(3cos$^2$θ -1)/2, where (ΔL/L)$_s$ is the saturation magnetostriction, and θ the angle between the magnetization and the easy axis [15]. The contrasting ΔL/L confirms a different easy axis in Region I and II. In **Figs.2x-2y**, these magnetostrictive effects are schematically illustrated but greatly exaggerated for clarity.

The magnetotransport exactingly follows the magnetostriction in an *anisotropic* manner (**Fig.3**). The magnetoresistivity $\rho_c(H\|a)$ closely tracks $\Delta a/a$ ($H\|a$ axis) and drops precipitously at H* (e.g., **Fig.3b**), whereas $\rho_c(H\|b)$ shows no such changes at H* (e.g., **Fig.3b**) but an expected drop at $H_c$ (**Figs.3b, 3d** and **3e**). With varying details, the data in **Fig.3** suggest two distinct, coexisting correlations: **(1)** When H $\|$ $a$ axis, $\rho_c(H\|a)$ strictly traces H* but shows no response to $H_c$, revealing a strong interlock between $\rho_c(H\|a)$ and the lattice, and **(2)** when H $\|$ $b$ axis, $\rho_c(H\|b)$ sharply responds to $H_c$ but not at all to H*, confirming a close link between $\rho_c(H\|b)$ and the spins. Note that this behavior is also true for $\rho_a$.

Indeed, the T-dependence of H* and $H_c$ is vastly different below 40 K (**Fig.4a**). H* decreases quickly with T whereas $H_c$ remains essentially unchanged below 40 K. This distinction underscores a dissociation between H* and $H_c$ as the former is related to the lattice and the latter the spins. They are activated by different field orientations, thus enabling the vastly different magnetotransport behaviors.

Considering the anisotropic magnetoelastic coupling, our investigation extends to include application of pressure, P, so that the transport behavior can be probed as functions of both P and H. The application of modest P rapidly suppresses $T_{MI}$ to 25.6 K at 20 kbar from 48 K at ambient pressure at a rate of $\Delta T_{MI}/\Delta P$ = -1.12 K/kbar, a result of the $t_{2g}$ band broadening owing to the compressed unit cell (**Fig.4b**). Remarkably, $T_N$ changes at a much slower rate of $\Delta T_N/\Delta P$ = - 0.37 K/kbar, shifting merely to 48.6 K at 20 kbar from 56 K at ambient pressure (**Inset** in **Fig.4b**). The unparallel response of $T_{MI}$ and $T_N$ to P further underlines the unconventional correlation between the transport and magnetic properties, in line with the fact that $T_{MI}$ and $T_N$ do not take place simultaneously (also recall $T_{MI}$ = 357 K and $T_N$ = 110 K in $Ca_2RuO_4$ [2, 31]).

While kept at P = 20 kbar, $\rho_a$ (T) is measured at H||$a$ and $b$ axis, respectively. At H||$a$ axis, both $\rho_a$ and $T_{MI}$ quickly decrease, and a metallic state emerges at 14 T, in which $\rho_a$ is remarkably linear with T below 60 K, a sign of a non-Fermi liquid (**Fig.4c**). Conversely, at H||$b$ axis, both $\rho_a$ and $T_{MI}$ rapidly rise, compared to those at H = 0. These opposing changes in $\rho_a$ and $T_{MI}$ confirm the anisotropic magnetostriction – H||$a$ axis complements the band broadening by effectively reducing the orthorhombicity or tilts of $RuO_6$ octahedra, resulting in a metallic state, whereas H||$b$ axis enhances the orthorhombicity, completely cancelling the effect of the pressure-induced band broadening, leading to a full recovery of the nonmetallic state (**Figs 4b-4c**).

This is also evidenced in $\rho_a$ (H) taken at 1.8 K and 20 kbar (**Fig.4d**). Application of P further amplifies the lattice effect of H* but reverses the spin influence of $H_c$ on $\rho_a$ (H). For H||$a$ axis, H* is approximately 17 T at ambient pressure (**Fig.1a**) and is now reduced to 7 T at P = 20 kbar, at which $\rho_a$(H||$a$) *decreases* by 87%. On the other hand, for H||$b$ axis, $\rho_a$(H||$b$) *increases* rapidly at $H_c$ by 48% at P = 20 kbar, starkly contrasting with the 10-fold drop at $H_c$ at ambient pressure (**Fig.1a**). This suggests that the spin configuration is significantly altered by P although the onset of $H_c$ remains essentially unshifted. The data in **Fig.4d** further illustrate two pairs of the distinct couplings: $\rho$(H||$a$) is coupled with H* and $\rho$(H||$b$) with $H_c$.

*Conclusion.* A phase diagram as functions of both P and H is illustrated in **Fig.5**. At H = 0, application of P readily suppresses $T_{MI}$ because of the band broadening. At P = 20 kbar, the added application of H leads to the non-Fermi liquid state when H||$a$ axis, which stretches the *a* axis, weakening the orthorhombicity, but to a reentrance of the nonmetallic state when H||$b$ axis, which stretches the *b* axis, strengthening the orthorhombicity -- The anisotropic magnetostriction enables a controlled switching between the two opposing states.

The strongly anisotropic magnetoelastic coupling uncovered in this work is activated by different field orientations, and enables two pairs of distinctive couplings, i.e., $\rho(H\|a)$ with the lattice and $\rho(H\|b)$ with the spins, thus the contrasting behavior of $\rho(H\|a)$ and $\rho(H\|b)$. These phenomena evoke the reported polar domain structures in $Ca_3Ru_2O_7$, in which a strong coupling between the polar moment and strain [12] could lead to the strong anisotropy of the magnetoelastic effect, mimicking a multi-order correlation in multiferroics, which merits more investigations. Nonetheless, this work offers much-needed insights into the intriguing physics of $Ca_3Ru_2O_7$ and, perhaps, a paradigm for an exquisite control of the electrical transport via anisotropy of flexible lattices in spin-orbit-coupled materials hosting a delicate interplay of essential interactions.

**Acknowledgement** This work is supported by NSF via grant DMR 1903888. GC is immensely thankful to Anshul Kogar and Feng Ye for their insightful comments on this work.

# References

1. Gang Cao and Lance E. De Long, Chapter 4, *Physics of Spin-Orbit-Coupled Oxides, Oxford University Press;* Oxford, 2021

2. G. Cao, S. McCall, M. Shepard, J.E. Crow and R.P. Guertin, *Magnetic and Transport Properties of Single Crystal $Ca_2RuO_4$: Relationship to Superconducting $Sr_2RuO_4$*, Phys. Rev. B **56**, R2916 (1997)

3. Y. Maeno, H. Hashimoto, K. Yoshida, *et al. Superconductivity in a layered perovskite without copper*, Nature **372,** 532–534 (1994)

4. G. Cao, S. McCall, J.E. Crow and R.P. Guertin*, Observation of a Metallic Antiferroamgnetic Phase and Metal-Nonmetal Transition in $Ca_3Ru_2O_7$*, Phys. Rev. Lett. **78**, 1751 (1997)

5. G. Cao, X.N. Lin, L. Balicas, S. Chikara, J.E. Crow and P. Schlottmann*, Orbitally-driven Behavior: Mott Transition, Quantum Oscillations and Colossal Magnetoresistance in Bilayered $Ca_3Ru_2O_7$, New Journal of Physics* **6**, 159 (2004)

6. G. Cao and P. Schlottmann*, $Ca_3Ru_2O_7$: A new paradigm for spintronics*, Modern Physics Letters B **22**, 1785-1813 (2008)

7. Gang Cao, Lance DeLong and Pedro Schlottmann*,* Chapter 6*, Frontiers of 4d- and 5d-Transition Metal Oxides*, Gang Cao and Lance E. De Long, ed. *World Scientific*, Singapore, 2013

8. D. J. Singh and S. Auluck, *Electronic Structure and Bulk Spin-Valve Behavior in $Ca_3Ru_2O_7$,* Phys. Rev. Lett. **96**, 097203 (2006)

9. V. Durairaj, X. N. Lin, Z.X. Zhou, S Chikara, E. Elhami, P. Schlottmann and G. Cao, *Observation of Quantum Oscillations Periodic in 1/B and B in Bilayered $Ca_3Ru_2O_7$*, Phys. Rev. B **73**, 054434 (2006)

10. X. N. Lin, Z.X. Zhou, V. Durairaj, P. Schlottmann and G. Cao, *Colossal Magnetoresistance by Avoiding a Ferromagnetic State in Mott System $Ca_3Ru_2O_7$*, *Phys. Rev. Lett.* **95**, 017203 (2005)

11. G. Cao, L. Balicas, Y. Xin, J.E. Crow, and C.S. Nelson, *Colossal Magnetoresistance, Quantum Oscillations and Magnetoelastic Interactions in Bilayered $Ca_3Ru_2O_7$*, *Phys. Rev. B* **67**, 184405 (2003)

12. Shiming Lei, Mingqiang Gu, Danilo Puggioni, Greg Stone, Jin Peng, Jianjian Ge, Yu Wang, Baoming Wang, Yakun Yuan, Ke Wang, Zhiqiang Mao, James M. Rondinelli and Venkatraman Gopalan, *Observation of Quasi-Two-Dimensional Polar Domains and Ferroelastic Switching in a Metal, $Ca_3Ru_2O_7$*, Nano Lett. **18**, 3088 (2018)

13. Note that recent studies on $Eu_5In_2Sb_6$ (npj Quantum Mater. **5**, 52 (2020)) and $Mn_3Si_2Te_6$ (PRB **103**, L161105 (2021)) report colossal magnetoresistance which, to a different extent, is associated with fluctuations.

14. S. McCall, G. Cao, and J.E. Crow, *Impact of Magnetic Fields on Anisotropy in $Ca_3Ru_2O_7$*, *Phys. Rev. B* **67**, 094427 (2003)

15. J. M. D. Coey, *Magnetism and Magnetic Materials*, *Cambridge University Press*, Cambridge, 2010

16. Y. Tokura, *Critical features of colossal magnetoresistive manganites*, *Rep. Prog. Phys.* **69**, 797 (2006)

17. A. Asamitsu, Y. Moritomo, Y. Tomloka, T. Arlmat & Y. Tokura, *A structural phase transition induced by an external magnetic field*, *Nature* **373**, 402 (1995)

18. M. B. Salamon, and M. Jaime, *The physics of manganites: Structure and transport*, *Rev. Mod. Phys.* **73**, 583 (2001)

19. M. M. Altarawneh, G.-W. Chern, N. Harrison, C. D. Batista, A. Uchida, M. Jaime, D. G. Rickel, S. A. Crooker, C. H. Mielke, J. B. Betts, J. F. Mitchell, and M. J. R. Hoch, *Cascade of Magnetic Field Induced Spin Transitions in LaCoO$_3$*, Phys. Rev. Lett. **109**, 037201 (2012)

20. V. V. Platonov, Y. B. Kudasov, M. P. Monakhov, and O. M. Tatsenko, *Magnetically induced phase transitions in LaCoO$_3$ in fields of up to 500 T*, Phys. Solid State **54,** 279–282 (2012)

21. Igor Markovic, Matthew D. Watson, Oliver J. Clark, *et al.*, *Electronically driven spin-reorientation transition of the correlated polar metal Ca$_3$Ru$_2$O$_7$,* PNAS **117***,* 15524 (2020)

22. There are some differences in transport properties between the flux- and floating-zone-grown Ca$_3$Ru$_2$O$_7$. This is largely because the flux-grown single crystals are grown in air whereas the floating-zone-grown single crystals are grown in an oxygen-rich environment. Ca$_3$Ru$_2$O$_7$ is known to be extremely sensitive to impurity doping including oxygen doping [1].

23. G. K. White and J. G. Collins, *Thermal expansion of copper, silver, and gold at low temperatures*, J. Low Temp. Phys. **7**, 43 (1972)

24. Magnetic properties were measured using a Quantum Design (QD) MPMS-7 SQUID Magnetometer. The measurements of electrical resistivity and magnetostriction were carried out using a QD Dynacool PPMS System with a 14-Tesla magnet. A hydrostatic pressure cell compatible with the QD PPMS was used for electrical resistivity as a function of pressure. Thermal expansion and magnetostriction were measured using a home-made dilatometer compatible with the QD 14T-PPMS-Dynacool system. The dilatometer is made with four identical KYOWA, type KFL strain gauges forming a Wheatstone bridge with the sample mounted on one arm and the rest three as compensators to cancel unwanted changes in the strain gauges due to changes in temperature and/or magnetic field. The thermal expansion results are qualitatively consistent with those independently collected from our sample at Quantum Design but the values of α are smaller than those of QD's

25. G. Cao, K. Abbound, S. McCall, J.E. Crow and R. P. Guertin, *Spin-Charge Coupling for Dilute La-doped $Ca_3Ru_2O_7$*, Phys. Rev. B **62**, 998 (2000)

26. Wei Bao, Z. Q. Mao, Z. Qu, and J. W. Lynn, *Spin Valve Effect and Magnetoresistivity in Single Crystalline $Ca_3Ru_2O_7$*, Phys. Rev. Lett. **100,** 247203 (2008)

27. J.F. Karpus, R. Gupta, Barath, S.L. Cooper and G. Cao, *Field-Induced Orbital and Magnetic Phases in $Ca_3Ru_2O_7$: Light Scattering Studies,* Phys. Rev. Lett. **93**, 167205 (2004)

28. H.L. Liu, S. Yoon, S.L. Cooper and G. Cao, *Raman Scattering Study of the Charge and Spin Dynamics of the Layered Ruthenium Oxide $Ca_3Ru_2O_7$*, Phys. Rev. B **60**, **R**6980, (1999)

29. C.S. Snow, S.L. Cooper, G. Cao, J.E. Crow, S. Nakatsuji and Y. Maeno, *Pressure-Tuned Collapse of the Mott-like State in $Ca_{n+1}Ru_nO_{3n+1}$ (n=1,2): Raman Spectroscopic Studies*, o, *Phys. Rev. Lett*. **89**, 226401 (2002)

30. E. Schierle, J. C. Lang, G. Srajer, S. I. Ikeda, Y. Yoshida, K. Iwata, S. Katano, N. Kikugawa, and B. Keimer, *Magnetic structure and orbital state of $Ca_3Ru_2O_7$ investigated by resonant x-ray diffraction*, Phys. Rev. B **77**, 224412 (2008)

31. C.S. Alexander, G. Cao, V. Dobrosavljevic, E. Lochner, S. McCall, J.E. Crow and P.R. Guertin, *Destruction of the Mott Insulating Ground State of $Ca_2RuO_4$ by a Structural Transition*, Phys. Rev. B **60**, R8422 (1999)

**Captions**

**Fig.1. (a)** The magnetic field dependence of magnetization for the *a*, *b* and *c* axis, $M_a$, $M_b$ and $M_c$ (solid lines, left scale) and the *c*-axis electrical resistivity $\rho_c(H\|a)$, $\rho_c(H\|b)$ and $\rho_c(H\|c)$ (dashed lines, right scale) [1]. Note that the *a*-axis resistivity $\rho_a$ shows the same behavior as $\rho_c$ does [10]. **(b)** The temperature dependence of the *a*, *b* and *c* axis [11] and **(c)** the thermal expansion coefficient α (blue and red lines, left scale) and the *a*-axis resistivity $\rho_a$ (black line, right scale). **Insets**: Single-crystal $Ca_3Ru_2O_7$ and distorted basal plane of the unit cell. Note that the shaded region in (b) highlights a rapidly changing *c* axis in the interval 40 K- 48 K.

**Fig.2.** The magnetic field dependence of magnetostriction ΔL/L **(a)-(f)** for ΔL/L(H∥*a* axis) and **(A)-(F)** for ΔL/L (H∥*b* axis) at six representative temperatures; **(x)-(y)** Schematic illustrations of the magnetostrictive effects on the lattice. Note that H* is marked by the black solid circles, and $H_c$ by the light gray dashed line.

**Fig.3. (a)-(e)** The magnetic field dependence of $\rho_c(H)/\rho_c(0)$ (solid lines, left scale) for H∥*a* (thick blue lines) and H∥*b* (thin red line) and ΔL/L(H∥L) (dashed lines, right scale) for comparison. Note that the black dashed line is the guide to the eye joining H*, and $\rho_c(H)/\rho_c(0)$ for H∥*b* is not displayed in (a) and (c) because it remains essentially the same as that in (b).

**Fig.4. (a)** The temperature dependence of H* and $H_c$. Note the two data points in green squares are estimated based on the resistivity [10]. The temperature dependence of the *a*-axis resistivity $\rho_a$ **(b)** at H = 0 and P = 1 bar, 5 kbar, 10 kbar, 15 kbar and 20 kbar, and **(c)** at P = 20 kbar and $\mu_0$H = 0 (black line), 7 T (thin lines) and 14 T (thick lines) for H∥*a* axis (blue lines) and H∥*b* axis (red

lines). **Inset**: $\rho_a$ at H = 0 and P = 20 kbar. **(d)** The magnetic field dependence of $\rho_a(H)/\rho_a(0)$ at P=20 kbar for H||*a* axis (blue line) and H||*b* axis (red line).

**Fig.5.** A phase diagram for $T_{MI}$ (thick lines) and $T_N$ (dashed thin lines) as functions of pressure and magnetic field for H||*a* axis (blue lines) and H||*b* axis (red lines) at P = 20 kbar. The schematic illustrations of the magnetostrictive effects on the orthorhombicity due to P and H. Note that PM stands for paramagnetic, M for metal and NM for nonmetal.

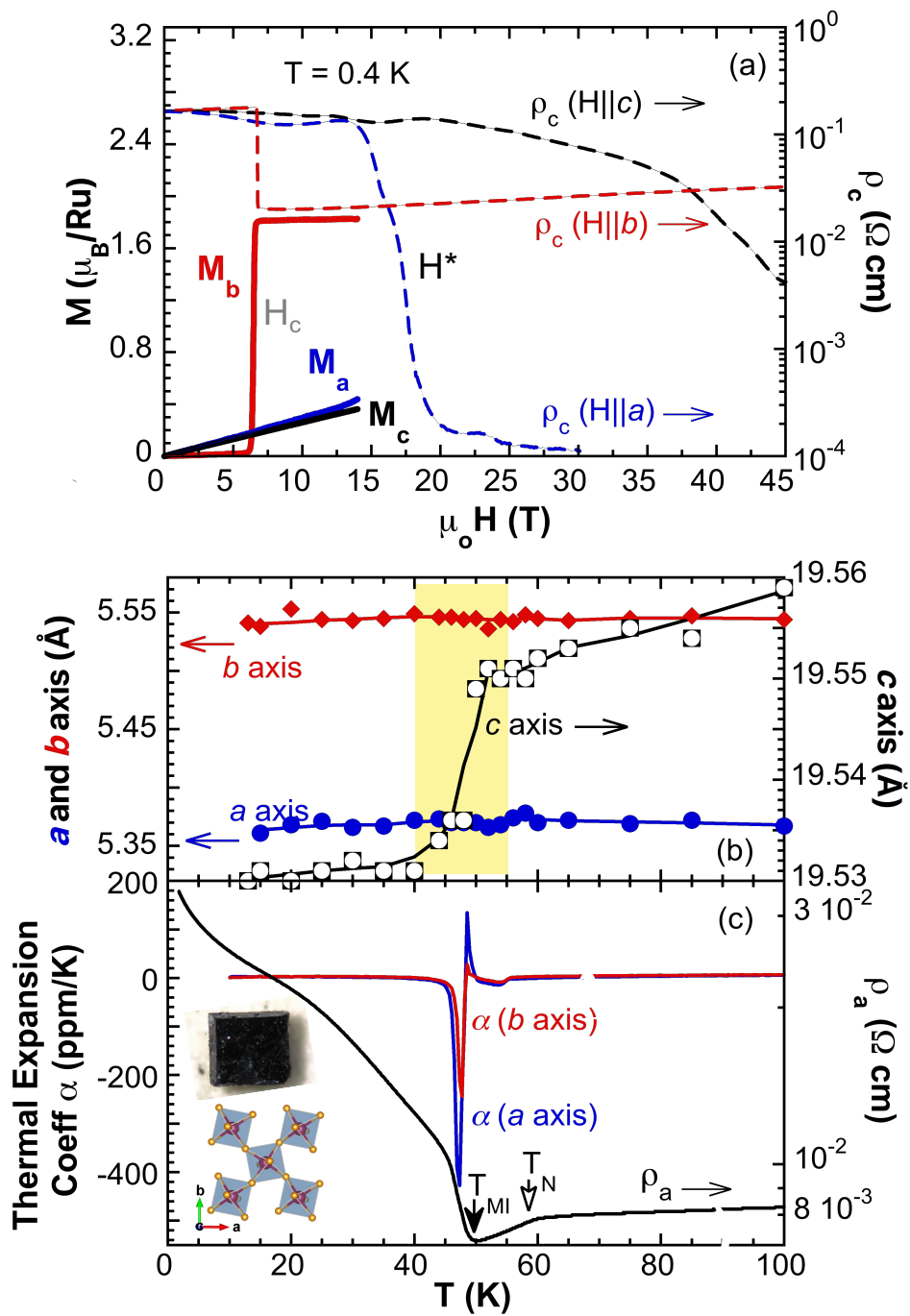

Fig. 1

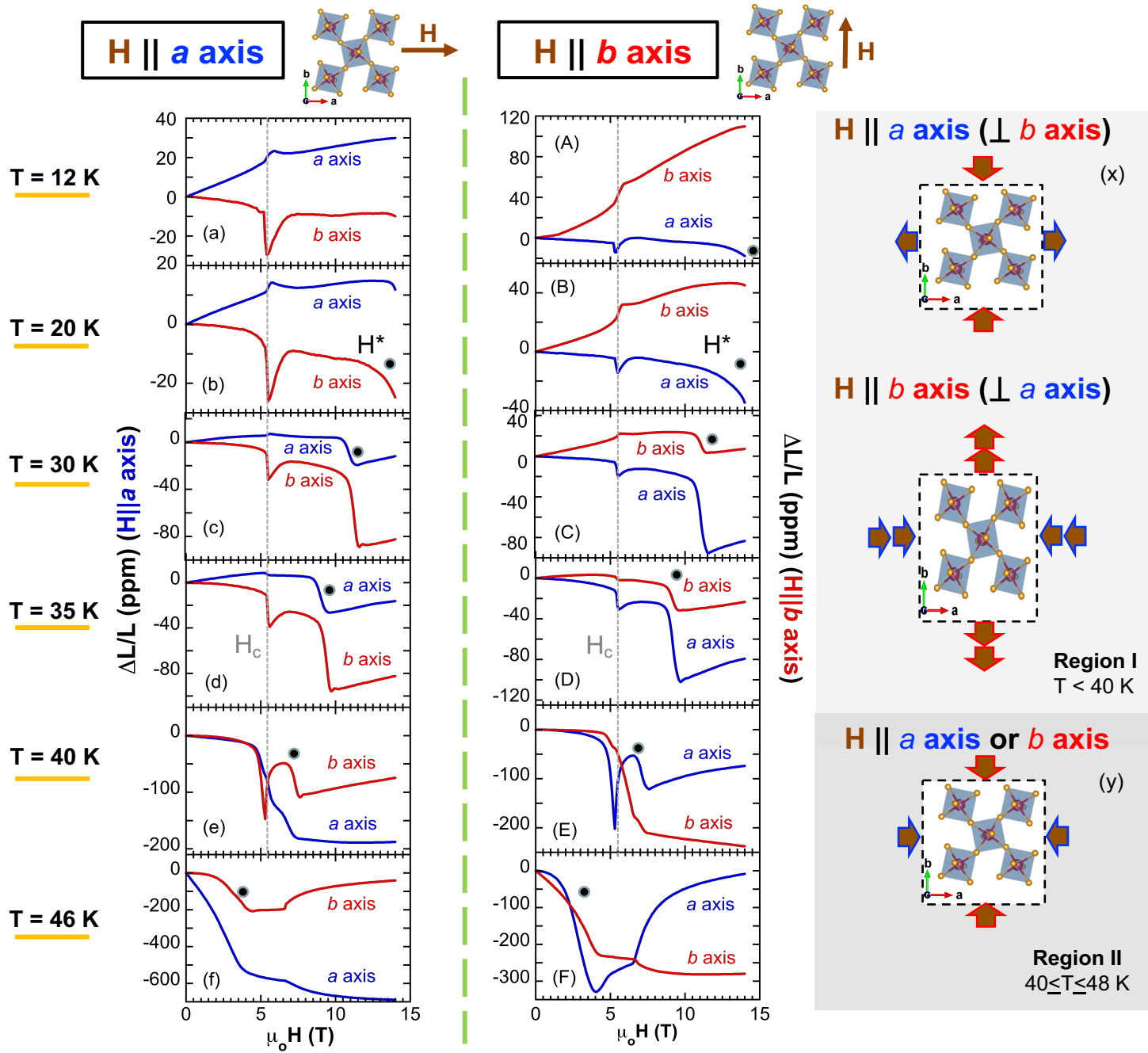

Fig. 2

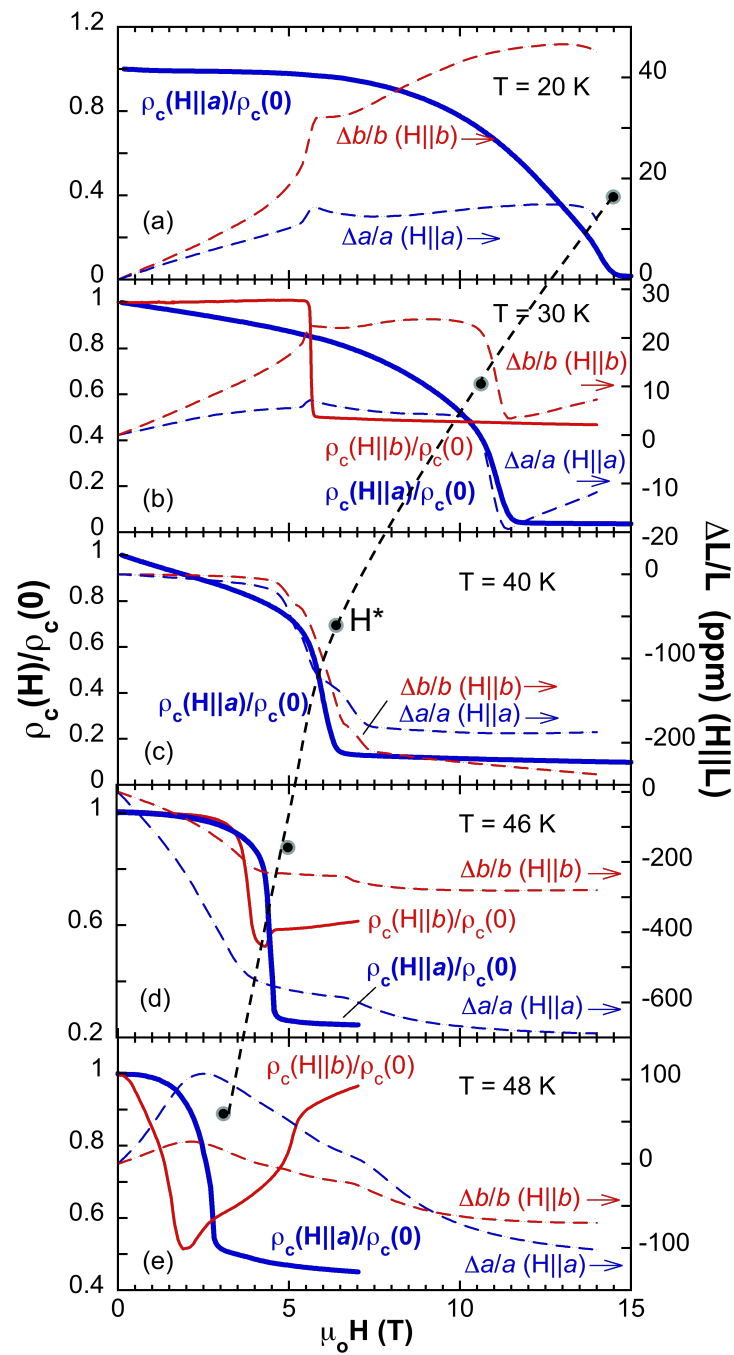

Fig. 3

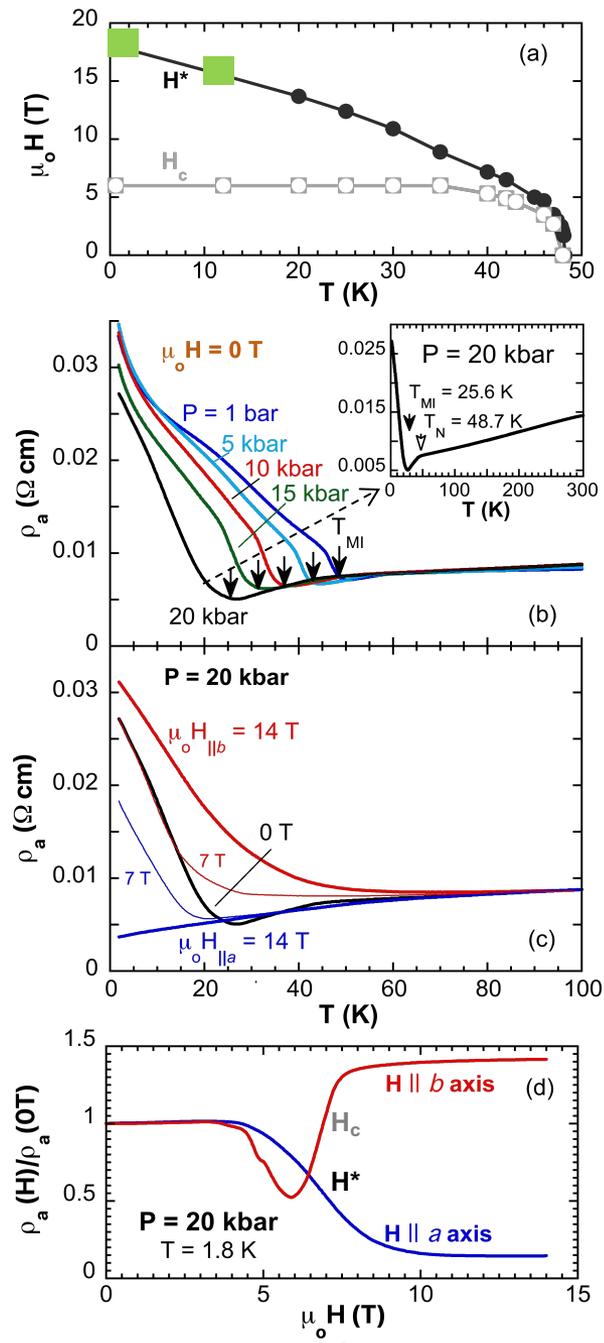

Fig. 4

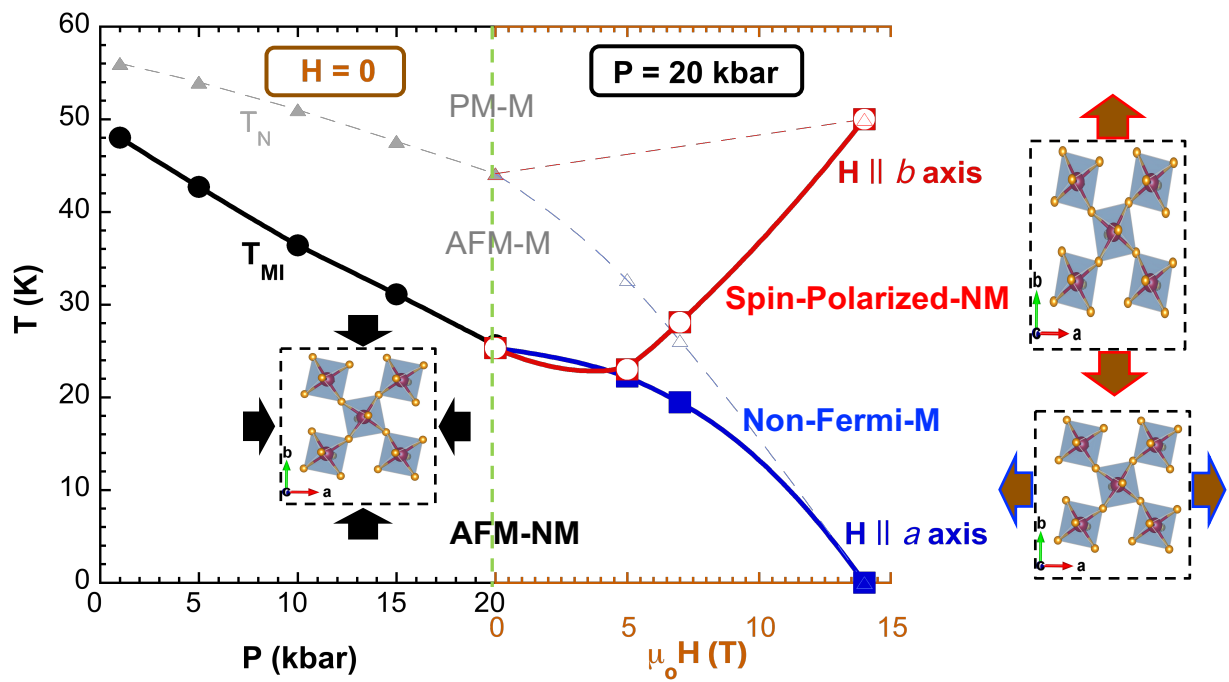

Fig. 5